\documentclass[aps,prl,
twocolumn,
showpacs,
superscriptaddress,
amsfonts,amssymb,amsmath,
tightening,
twoside,
letterpaper,
nobibnotes,
]{revtex4}

\usepackage{bm,color,mathrsfs,graphicx,hyperref}
\usepackage{cancel} 

\hypersetup{
    colorlinks=true, 
    linkcolor=blue,  
    citecolor=blue,  
}

\begin{document}

\title{
Single- and multi-domain ferroelectricity driven by interfaces
}

\author{A. Cano}
\affiliation{
CNRS, Univ. Bordeaux, ICMCB, UPR 9048, F-33600 Pessac, France
}
\affiliation{
European Synchrotron Radiation Facility, 6 rue Jules Horowitz, BP 220, 38043 Grenoble, France
}
\author{A.P. Levanyuk}
\affiliation{
\mbox{
Department of Physics, University of Washington, Seattle, Washington 98195, USA
}
}
\affiliation{
Moscow State Technical University of Radioengineering, Electronics and Automation (MSTU-MIREA),
Prospect Vernadskogo 78, Moscow 119454, Russia
}

\date{\today}

\begin{abstract}
The design of the interfacial bondings at metal-oxide interfaces yields exciting new phenomena and can be a route to sustain, and even promote, ferroelectricity at the nanoscale. We study the impact of these interfaces on the nature of the spontaneous polarization (single- vs. multi-domain) of ferroelectric capacitors. 
We show that interfacial properties interwine with both ferrolectric and electrode parameters to determine the actual ground state of the system.
We find analytically the criterion that specifies if ferroelectricity appears in a single- or multi- domain fashion as a result of this intertwining. The physics behind this criterion suggests new means for tailoring ferroelectric functionalities.
\end{abstract}

\pacs{
}

\maketitle

At present, there is a considerable ongoing effort in fabricating nanoscale ferroelectric devices suitable for memory and other technological applications \cite{Fong04,RAT}. 
In these systems, several factors, both intrinsic and extrinsic, are known to have a substantial impact on the ferroelectric transition. Its temperature, for example, strongly depends on the amount of free charges available to screen the depolarizing field created at the interfaces 
\cite{Ivanchik62,
Junquera08,
Bratkovsky09}. 
Beyond that, the nature of the transition itself is a subject of question as ferroelectricity can appear in either single- or multi-domain fashion depending on various additional factors \cite{Bratkovsky06,Aguado08,Bratkovsky09}. 

\begin{figure}[tb]
\includegraphics[width=.425\textwidth]{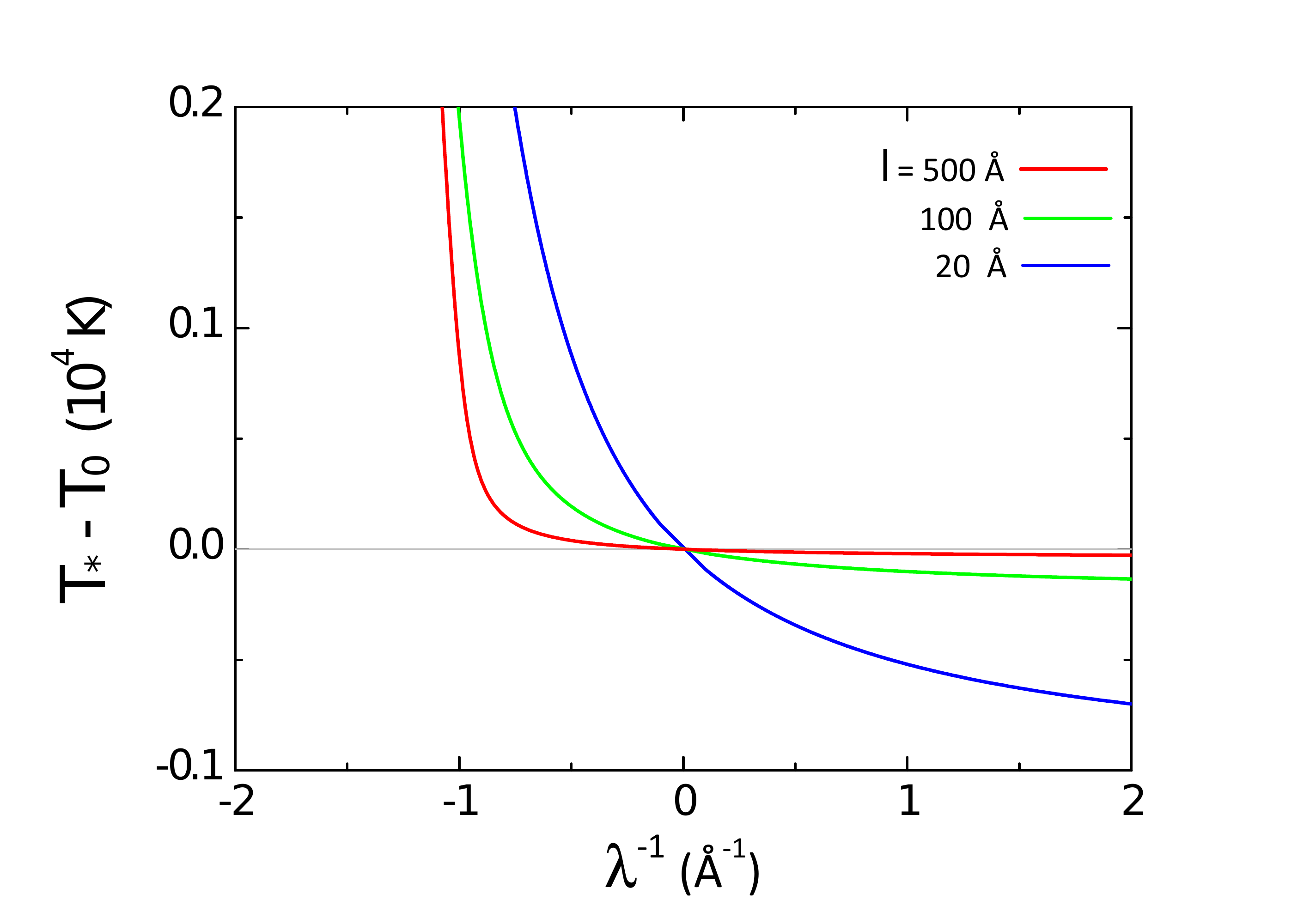}
\put(-70,100){\includegraphics[width=.125\textwidth]{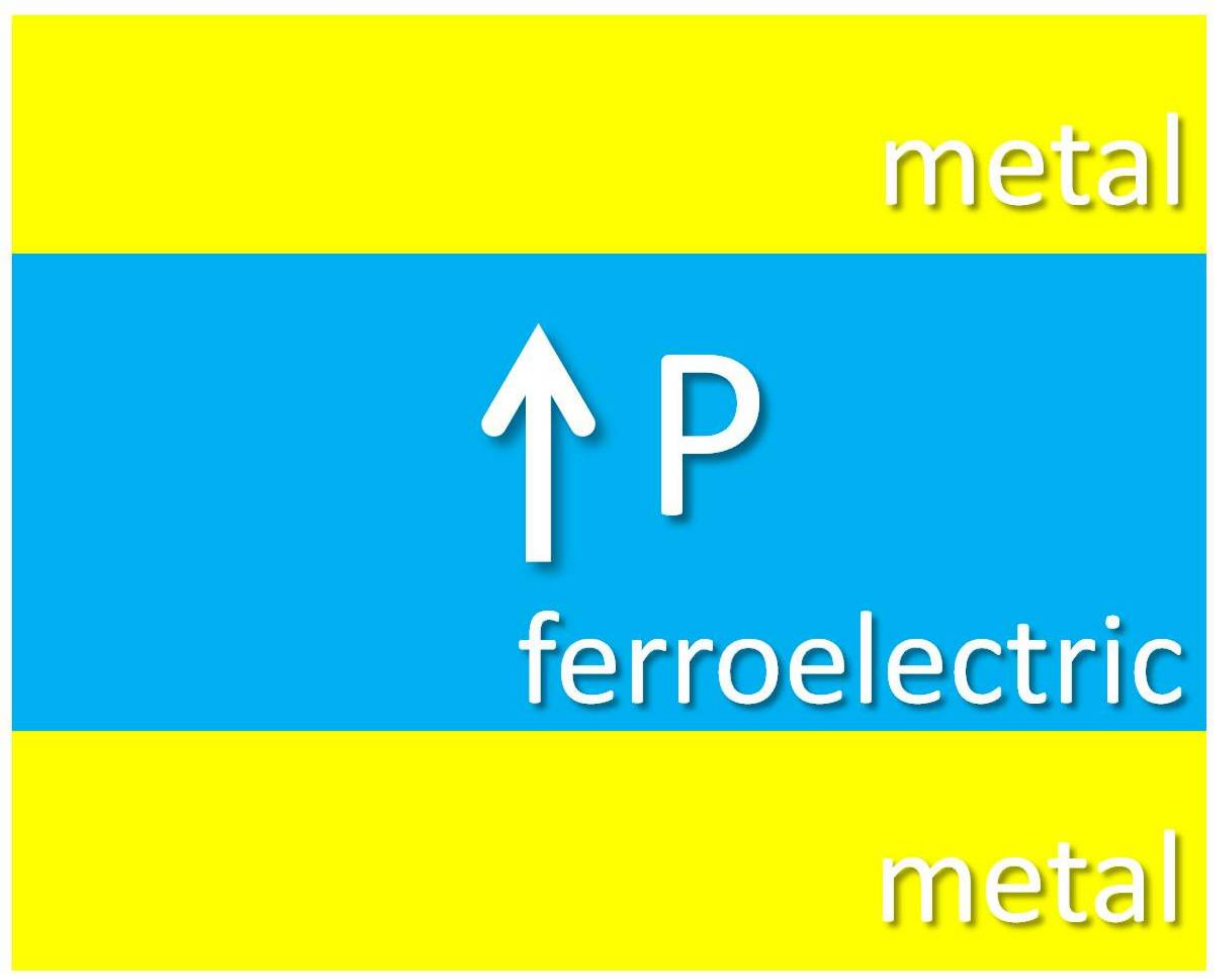}}
\put(-180,52){
$l = 500 $ \AA \; $^\text{\textcolor{red}{\thicklines\line(1,0){20}}}$
}
\put(-180,42){
\; \; \, $100 $ \AA \; $^\text{\textcolor{green}{\thicklines\line(1,0){20}}}$
}
\put(-180,32){
\; \; \, \, $20 $ \AA \; $^\text{\textcolor{blue}{\thicklines\line(1,0){20}}}$
}
\caption{Relative transition temperature $T_c^{(0)}-T_0$ for the single-domain state as a function of the inverse of the extrapolation length $\lambda$ in a ferroelectric capacitor with ideal electrodes. $\lambda^{-1} = 0 $ corresponds to passive interfaces such that $T_c^{(0)}$ coincides with the ``bulk'' transition temperature $T_0$ (one then deals with the so-called natural boundary conditions $\partial_z P =0$). For $-1<\lambda^{-1} \leq 0$ ferroelectricity is enhanced at the interfaces and $T_c^{(0)} > T_0$ while for $\lambda^{-1} \geq  0$ it is suppressed and $T_c^{(0)} < T_0$.}
\label{F1}
\end{figure}

Stengel {\it et al.} have shown from first principles calculations that the right choice of the metal-ferroelectric interface can notably enhance ferroelectricity due to bonding effects \cite{Stengel09}. This observation is very appealing, as it opens new routes in nanoscale device designing. In terms of the Landau theory these bonding effects are described by means of interfacial free energy contributions as first discussed in \cite{Kaganov71,Kretschmer79}. A negative interface energy enhances the overall ferroelectricity as confirmed in \cite{Stengel09} and, in any case, dissimilarities between bulk and interface properties generate space variations of the electric polarization as described in \cite{Kretschmer79}. 
These variations, though confined in narrow regions near the interfaces, yield a dependence of the transition temperature on the thickness of the ferroelectric, even if the screening of the electrodes is perfect as illustrated in Fig. \ref{F1}. The reason is that the (volume) bound charges created in these regions remain unbalanced, and therefore generate a depolarizing field. 
In this paper we show that this intrinsic depolarizing field has a far more drastic influence on the ferroelectric transition, 
as it suffices to boost the appearance of multi-domain structures that spoil single-domain ferroelectricity. 
This unexpected link between surface energy effects and the appearance of multi-domain structures has been noticed by Morozovska {\it et al.} for the special case of a ferroelectric with a type-II incommensurate phase from phase-field numerical simulations \cite{morozovska10}. 
In the following we demonstrate the generality of this this novel depolarizing-field mechanism and determine analytically the conditions under which the ferroelectric single-domain state is protected against it.

For this, we develop a generalization of the Landau-like approach that incorporates interfacial bonding effects into the description of the paraelectric instability in thin films. 
This approach is especially conceived to capture fundamental physics near phase transitions and, although expected to be valid only at macroscopic scales, has repeatedly been successful in describing most of the properties of ferroelectric thin films and heterostructures down to the nanoscale \cite{Bratkovsky09,Tagantsev06,Chandra07}.
In this way, we show that interfacial bonding effects intervene conjointly with both ferroelectric and electrode properties in a non-trivial way. 
We find, in particular, a criterion in terms of ferroelectric stiffness, electrode screening, and interface energy that specifies whether ferroelectricity appears in single or multi-domain fashion [see Eq. \eqref{criterion} below]. Remarkably, there is a ferroelectric anisotropy factor that largely determines this criterion. This factor is inherently large for cubic ferroelectrics like BaTiO$_3$ that become tetragonal due to substrate misfit. In fact there exists an anisotropy threshold beyond which, even if the electrode screening is perfect, the single-domain state is not preserved unless the interfaces are completely passive. This unanticipated interplay is indeed a crucial point when it comes to the design of ferroelectric devices with enhanced functionalities. 

We consider uniaxial ferroelectrics or cubic systems that become effectively uniaxial as we said before (in BaTiO$_3$/SrRuO$_3$/SrTiO$_3$ structures for example). 
BaTiO$_3$ and PbTiO$_3$, in particular, represent model systems for the physics we discuss in following because of i) its versatility as regards interfaces \cite{Stengel09} and ii) its sizeable anisotropies \cite{Hlinka06}. For a parallel-plate ferroelectric capacitor, with the spontaneous polarization $P$ perpendicular to the plates (see inset in Fig. \ref{F1}), the problem is formulated as follows. The ferroelectric response is described by the constitutive equation that follows from the corresponding Landau free energy functional. In the capacitor setup this functional contains bulk and interface contributions:
\begin{align}
F= F_\text{bulk} + F_\text{interface}.  
\label{F}\end{align}
The ferroelectric instability implies a diverging electric susceptibility which, for a complete analysis, makes it necessary to consider spatial derivatives of the ferroelectric polarization $P$ and non-linear terms. Following the Landau approach $F$ is expanded in powers of $P$ and its derivatives, and the constitutive equation is obtained as ${\delta F\over \delta P} = E_z$, where $\mathbf E$ is the electric field (the ferroelectric axis is taken as the $z$-axis hereafter). 
To obtain i) the transition temperature and ii) the form (but not the amplitude) of the distribution of polarization that appears below the transition, this equation can be written as
\begin{align}
[A - C_\perp (\partial_x^2 +\partial_y^2 )- C_z \partial_z^2]P= E_z,
\label{constitutive}\end{align}
where $A = A'(T-T_0)$ represents the nominal inverse susceptibility, and $C_z$ and $C_\perp$ are expansion coefficients that determine the stiffness of the polarization with respect to space variations (parallel and perpendicular to the ferroelectric axis respectively). 
The electrostatics of the problem is governed by the Gauss's law, $\nabla \cdot (\mathbf E + \varepsilon_0^{-1}\mathbf P)=0$, and the Maxwell-Faraday equation, $\nabla \times \mathbf E = 0$. The later is automatically satisfied by expressing the electric field as the gradient of the electric potential: ${\mathbf E}= - \nabla V$, while the former can be written as
\begin{align}
\varepsilon_0\varepsilon_\perp (\partial_x^2 +\partial_y^2 )V + \varepsilon_0\varepsilon_b \partial_z^2 V  - \partial_z P = 0,
\label{Gauss}\end{align}
where $\varepsilon_\perp $ is the relative permittivity perpendicular to the ferroelectric axis and $\varepsilon_b$ the ``background'' permittivity --due to extra (non-critical) contributions to the total polarization \cite{Tagantsev06}. Eqs. \eqref{constitutive} and \eqref{Gauss} describe the ``bulk'' behavior of the ferroelectric capacitor. 
In the case of BaTiO$_3$/SrRuO$_3$/SrTiO$_3$, in particular, 
$T_0 = 1273$K (``bulk''), 
$A'= 3\times 10^5 \rm J \cdot m \cdot C^{-2} \cdot  K^{-1}$, $C_z = 5.1 \times 10^{-10} \rm J \cdot m^3 \cdot C^{-2}$ and $C_\perp = 0.2\times 10^{-10} \rm J \cdot m^3 \cdot C^{-2}$, while $\varepsilon_\perp $ and $\varepsilon_b $ can reach values of $218$ and $7.35$ respectively \cite{Pertsev98,Bratkovsky09,Hlinka06}.

Finite size effects are conveniently described by means of boundary conditions supplementing the above equations. On one hand we have the electrostatic boundary conditions that, for a short-circuited capacitor with ideal electrodes, read
\begin{align}
V(z = \pm l/2) =0,
\label{EBC}\end{align}
where $l$ is the thickness of the ferroelectric. On the other hand, we have the additional boundary conditions for the ferroelectric polarization \cite{Kretschmer79}:
\begin{align}
\left.(1 \pm \lambda \partial_z ) P\right|_{z = \pm l/2} = 0. 
\label{ABC}\end{align}
This represents the boundary conditions for the constitutive equation Eq. \eqref{constitutive} 
in which the interface contribution $F_\text{interface} = {C_z\over 2\lambda }[P^2(-l/2)+P^2(l/2)]$ to the total free energy [see Eq. \eqref{F}] has been taken into account \cite{note_polar}. 
The quantity $\lambda $ is the so-called extrapolation length, which describes the difference between the properties of the ferroelectric in the bulk and at the interfaces. 
The typical values considered in the literature are $\sim 0.2-50$ nm \cite{Duan06,jia07,morozovska10}. 

\begin{figure*}[tb]
\includegraphics[width=.425\textwidth]{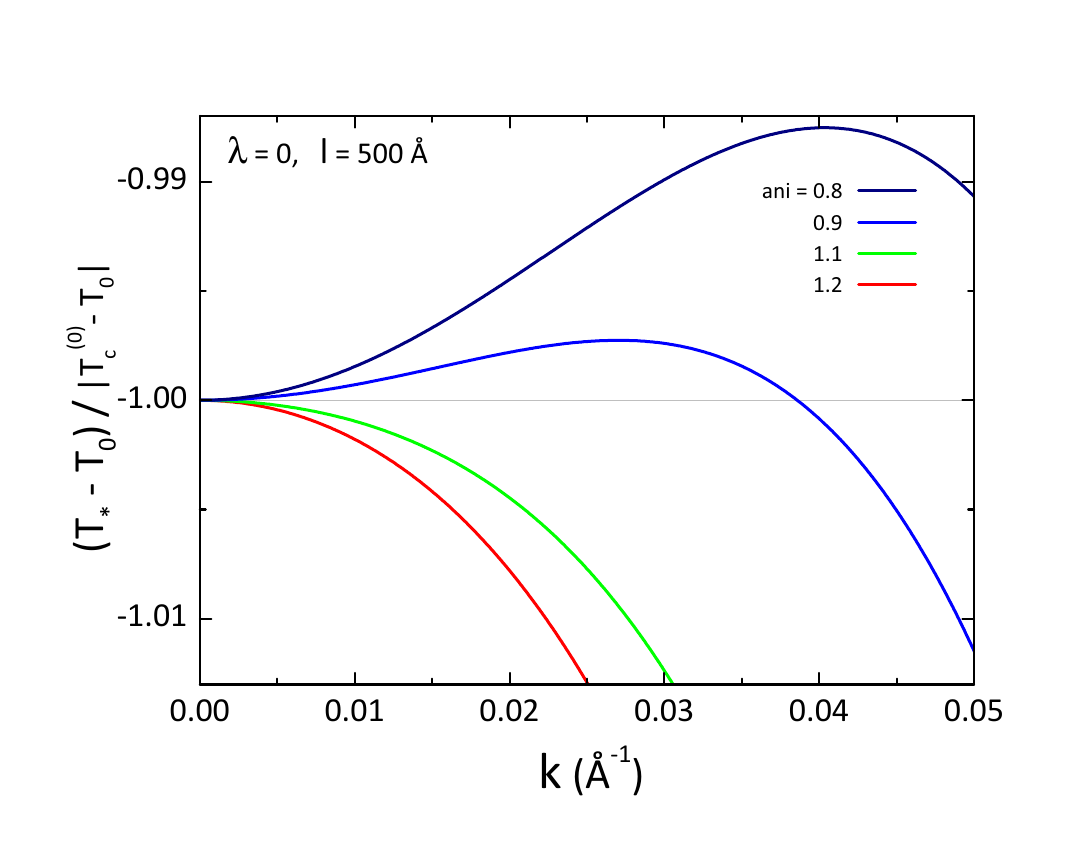}
\hspace{.05\textwidth}
\includegraphics[width=.425\textwidth]{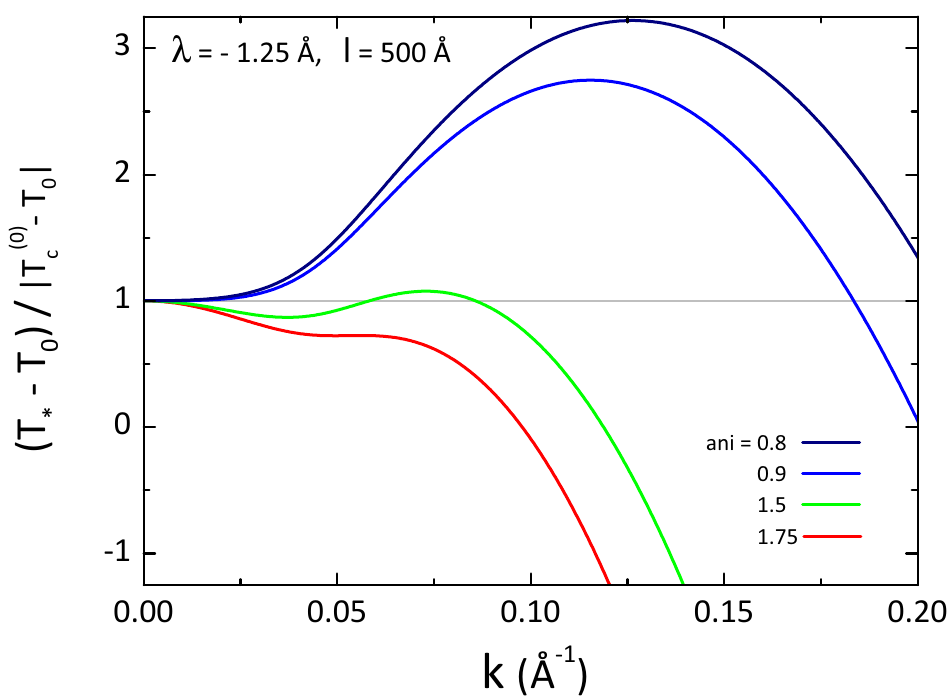}
\caption{Relative temperature $T_* - T_0$ at which the parametrization of the electric polarization described in the text gives a nontrivial solution of the set of equations as a function of the wavenumber for different values of the gradient anisotropy. (a) interface suppressed ferroelectricity ($\lambda \geq 0$) and (b) interface enhanced ferroelectricity ($-1<\lambda < 0$). The anisotropy is quantified by the parameter $\text{ani}= {3{(\varepsilon _b /\varepsilon _\perp) }(1+\kappa \lambda)^2} {(C_\perp/C_z)} $. For clarity, the curves have been normalized to the absolute value of the relative transition temperature $|T_c^{(0)} - T_0|$ expected for the single-domain state. The maximum of these curves gives the actual transition temperature and the period of the spontaneous polarization. }
\label{F2}
\end{figure*}

Our main objective is to answer the question of whether the intrinsic depolarizing field that gives rise to the behavior shown in Fig \ref{F1} can in fact change the nature of the ferroelectric transition from single-domain to multi-domain. For positive interface energy ($\lambda \geq 0$) this possibility is rather natural. The (intrinsic) depolarizing field concentrated near the interfaces can be seen as analogous to the (stray) depolarizing field that appears when the electrode screening is not perfect and/or there are dead layers at the interfaces. This hampers the single-domain state and its virtual appearance is pushed down in temperature as shown Fig. \ref{F1}. Then there appears a phase region between the ``bulk'' instability towards single-domain ferroelectricity at $T_0$ and the actual one that in principle is available for multi-domain structures (for which the concomitant depolarizing field can be considerably lower). Whether this phase space is finally taken, and the actual instability implies multi-domain ferroelectricity, is eventually determined by the anisotropy of the ferroelectric as we show below. 
This possibility is far less evident when the interface energy is negative ($\lambda < 0$). In this case the appearance of single-domain state is pushed up in temperature (see Fig. \ref{F1}), and therefore it seems that no phase-space is available for multi-domain structures. As we show below, these structures may still appear because, compared to the single-domain state, the interface enhancement pushes them to even higher temperatures if the anisotropy factor of the ferroelectric is sufficiently large.

This can be readily illustrated for the extreme case of a ferroelectric such that $C_\perp = 0$ and $|\lambda| \ll l$.
In this case, the modulations of the polarization $\perp$ to the film can be created at no energy cost. Thus their period can be arbitrarily small and the multi-domain transition is simply determined from the condition of absence of depolarizing field. We then can borrow the the formulas obtained in \cite{Kaganov71}. 
The single-domain transition is the expected to occur at $T_c^{(0)} \sim T_0 - 2 C_z / (A'\lambda l) $, while for the multi-domain transition we have $T_c^{(m)} \sim T_0 -\pi^2 C_z / (A'l^2) $ for $\lambda \geq 0$ and $T_c^{(m)} \sim T_0 + C_z / (A'\lambda^2) $ for $\lambda < 0$. As we see, the instability towards multi-domain ferroelectricity is reached much before the single-domain one if $\lambda \ll l$. In the following we confirm this expectation and describe the situation in the most general case (where $C_\perp \not = 0$).

We extend the approach developed in Ref. \cite{Chensky82} in order to incorporate interfacial bonding effects into the description of the paraelectric instability and the subsequent appearance of ferroelectricity (see Supplementary Material). Within this approach, one first focuses on a given period $2\pi/k$ of the distribution of polarization and then the general solution of the resulting problem reduces to $P = (p_1 \cos k_1 z + p_2 \cos k_2 z )\cos k x $ and $V = \big(p_1 {k_1 \over \varepsilon_0 \varepsilon_\perp k^2 + \varepsilon_0 k_1^2}
\sin k_1 z + p_2 {k_2 \over \varepsilon_0 \varepsilon_\perp k^2 + \varepsilon_0 k_2^2}\sin k_2 z \big)\cos k x $ for the symmetry of our setup, where the parameters of these expressions are determined self-consistently from Eqs. \eqref{constitutive}, \eqref{Gauss}, \eqref{EBC} and \eqref{ABC}. Thus the task is to find out the parameter $k$ that gives the solution for the highest temperature. Note that both electrostatic and bonding effects are taken into account simultaneously through the boundary conditions. As a result, we find that the instability necessarily implies multi-domain ferroelectricity ($k\not =0$) at $T_{c}^{(m)}$ if 
\begin{align}
3(1 + \kappa \lambda)^2{\varepsilon_b C_\perp \over \varepsilon_\perp C_z}< 1
\label{condition-on-gradients}\end{align}
while otherwise single-domain ferroelectricity ($k =0$) can step up at $T_c^{(0)}$. 
Here $T_{c}^{(0)} $ and $\kappa $ are related via $\kappa = [(1+ \varepsilon_0 A_0) /(\varepsilon_0 C_z)]^{1/2}$ and $A_0=-2C_z \kappa^2/[(1+\kappa \lambda)\kappa l - 2]$, where $A_0 = A'(T_c^{(0)}-T_0)$. 
To the best of our knowledge, the first possibility has been unnoticed so far as single-domain ferroelectricity is repeatedly taken for granted (with the exception of \cite{morozovska10}). In the following, we discuss the overall situation in detail. 

Fig. \ref{F2} shows the temperature $T_*$ at which the above expressions give nontrivial solutions as a function of their wavenumber $k$ for different anisotropies of the ferroelectric. The maximum of these curves gives the actual transition temperature. That is, the highest temperature at which a nontrivial solution is possible and consequently the paraelectric phase gets unstable. The condition \eqref{condition-on-gradients} is related to whether these curves go upwards or downwards in the limit $k\to 0$. If the curve goes upwards, this immediately tells us that the maximum is at $k\not = 0$ and then the transition necessarily implies multi-domain ferroelectricity. However, when the curve goes downwards, the condition \eqref{condition-on-gradients} is less conclusive due to the possible appearance of additional maxima at $k\not = 0$ (see below). Fig. \ref{F2} (a) corresponds to $\lambda = 0$, which is a representative case for ferroelectricity suppressed at the interfaces. Note that these curves have only one maximum and, accordingly, the transition implies a finite period that can vary continuously and eventually reach infinite by tuning the material parameters in such a way that Eq. \eqref{condition-on-gradients} is not fulfilled. For the interfaces enhancing ferroelectricity ($\lambda < 0$), however, we obtain curves that develop two maxima as can be seen in Fig. \ref{F2} (b). In this case, even if the condition \eqref{condition-on-gradients} is not fulfilled, multi-domain transition cannot be excluded because of the additional maximum. In fact, the position of the absolute maximum can change abruptly as it does with the anisotropy factor in Fig. \ref{F2} (b). When the modulation of the polarization reaches atomic distances, our results have to be understood at the qualitative level only and microscopic considerations are required. 

\begin{figure*}[tb]
\includegraphics[width=.425\textwidth]{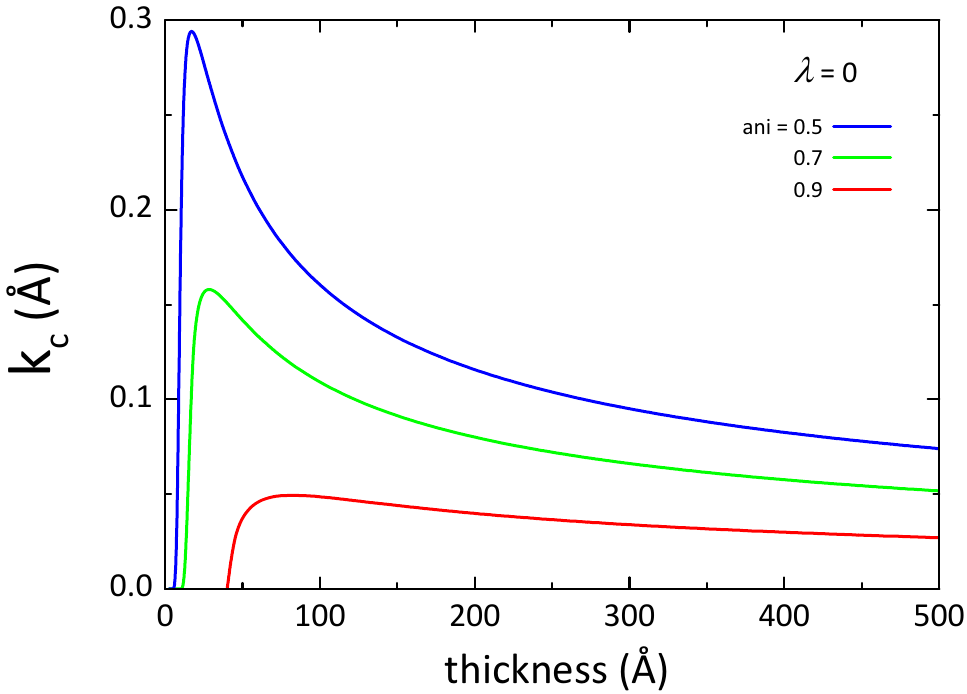}
\hspace{.05\textwidth}
\includegraphics[width=.425\textwidth]{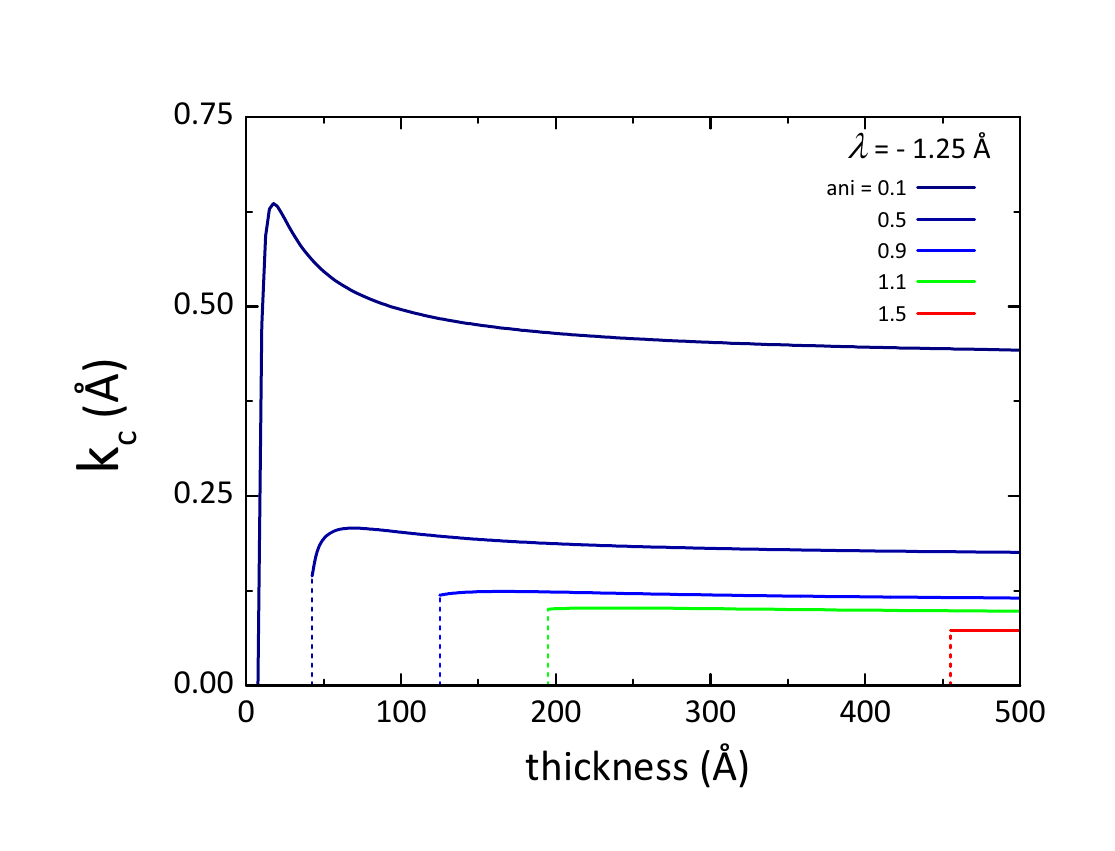}
\caption{Wavenumber $k_c$ of the distribution of polarization as a function of the thickness for a parallel-plate ferroelectric capacitor with ideal electrodes and (a) interfaces suppressing ferroelectricity ($\lambda \geq 0$), (b) interfaces enhancing ferroelectricity ($\lambda < 0$). The parameter $\text{ani}= {3{(\varepsilon _b /\varepsilon _\perp) }(1+\kappa \lambda)^2} {C_\perp/C_z} $ quantifies the gradient anisotropy. With decreasing the thickness, the wavenumber first increases and then decreases, eventually reaching the value $k_c = 0$ corresponding to single-domain ferroelectricity. This evolution is continuous in the case of interface enhanced ferroelectrictiy (a) while it can be either continuous or discontinuous for interface suppressed ferroelectricity (b). The latter is eventually determined by microscopic details.}
\label{F3}
\end{figure*}

The wavenumber $k_c$ of the critical distribution of polarization is shown in Fig. \ref{F3} as a function of the thickness. By decreasing the thickness, $k_c$ first increases and then decreases eventually reaching zero, which describes the single-domain state. This change is continuous if the interfaces hamper ferroelectricity [Fig. \ref{F3}(a)], while it can be discontinuous in the case of ferroelectricity enhanced at the interfaces [Fig. \ref{F3}(b)]. In the former case, the initial increase is strikingly similar to that of ferroelectric capacitors with poor electrode screening (which, on the contrary, we here consider as perfect). 
In fact, it follows the ``Kittel law'' $l^{1/2}$ as it is driven by the advantage of forming domains due to the corresponding reduction of the depolarizing field (which, in our case, is created by just the longitudinal variations of $P$ necessary to satisfy the boundary conditions).  
This advantage, however, is lost when the region in which this field is confined becomes comparable to the film thickness itself, and then the system tends to reach the single-domain situation (more favorable from the point of view of gradient energy). This drives the decrease of $k_c$ shown Fig. \ref{F3} in the ultrathin limit. The resulting crossover is similar to that reported 
in \cite{Cano10} for the case of multiferroics. 
Our results indicate that interplay between interfacial effects and anisotropy in the gradient stiffness suffices to promote this crossover from atomic distances to thicknesses of experimental relevance. We note that the thickness below which single-domain ferroelectricity is protected also depend on this interplay. 

\begin{figure}[b]
\includegraphics[width=.425\textwidth]{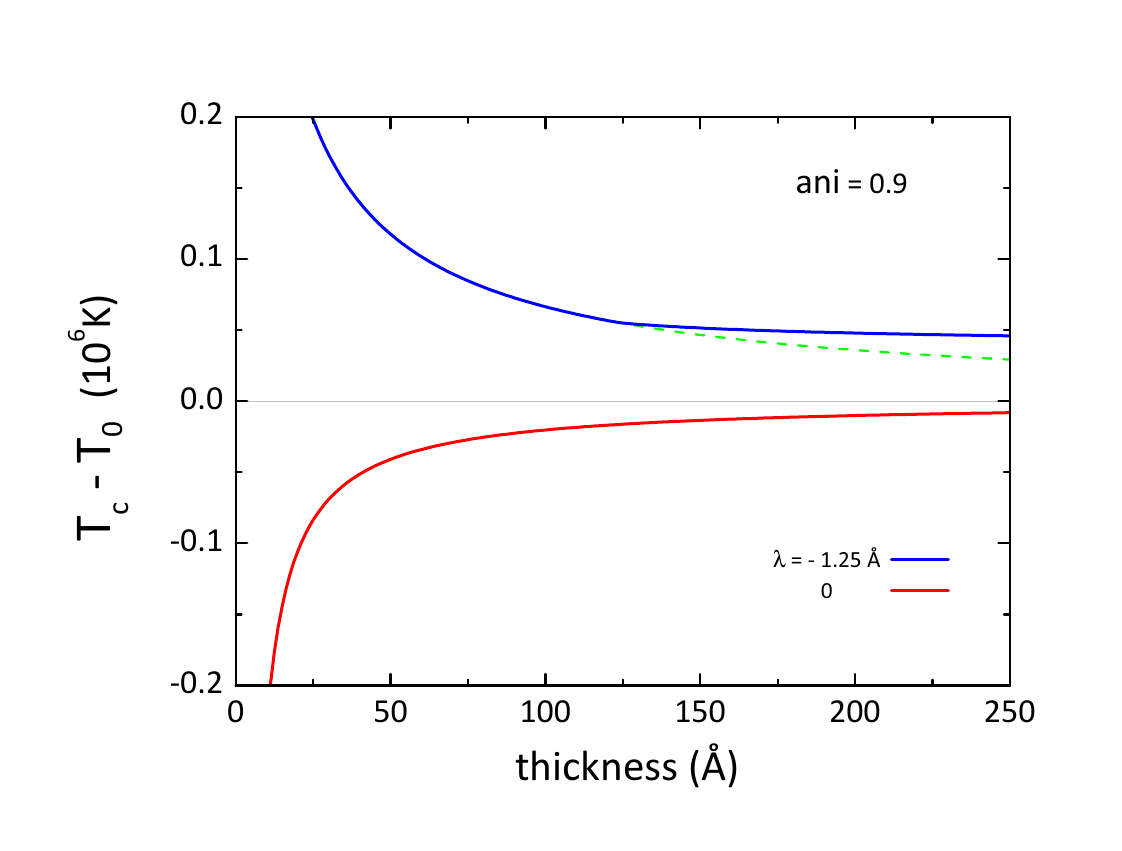}
\caption{Relative transition temperature $T_c - T_0$ as a function of the ferroelectric thickness for a capacitor with ideal electrodes. The gradient anisotropy is such that $\text{ani}= {3{(\varepsilon _b /\varepsilon _\perp) }(1+\kappa \lambda)^2} {C_\perp/C_z} =0.9$. 
In the case of interface enhanced ferroelectricity (red curve) the transition temperature increases with decreasing the thickness while it decreases in the case of interface suppression (blue curve). 
}
\label{F4}
\end{figure}

In Fig. \ref{F4} we plot the resulting transition temperature as a function of the film thickness. Both types of interfaces, enhancing and suppressing ferroelectricity, are considered. It is worth mentioning that in the plot for $\lambda = 0$ the transition has a multi-domain character, but for $\lambda < 0$ this changes abruptly to single-domain below $\sim$125 {\AA}. As we have mentioned. We also note that for $\lambda < 0$ the asymptotic behavior in the bulk limit ($l \to \infty $) does not converge to the nominal transition temperature $T_0$. This is because the unstable regions of the system are localized near the interfaces, which effectively decouple if the film is sufficiently thick. This means that the overall transition i) it is largely dominated by the interface instability (rather than the bulk one) and ii) is not affected by the total size of the system. Then the depolarizing field becomes relatively unimportant, and the situation is analogous to that described in \cite{Kaganov71}. 

We have demonstrated that the same interfacial boding effects that can enhance/suppress ferroelectricity can cause the appearance of multi-domain states in ferroelectric capacitors with nominally perfect electrodes. This implies that the actual screening in the device is entangled with the interfacial properties and therefore cannot be determined independently from the intrinsic properties of the metals (such as the corresponding Thomas-Fermi screening length). 
This is in tune with generic ideas put forward when discussing the so-called critical thickness for (single-domain) ferroelectricity \cite{Junquera08}. 
The critical screening strength necessary to keep the device in such a single-domain state, for example, has to be computed jointly. 
Thus, within a semi-classical approximation, in which the electric potential in the electrodes is $\sim e^{-|z|/\ell_s}$, the condition \eqref{condition-on-gradients} becomes
\begin{align}
{
3 (1+\kappa \lambda)^2
[1+(\varepsilon_{b} /\varepsilon_{m} )\kappa \ell_s(1 + \kappa \lambda)]
\over 
\big[
1 
+ { 2 (\varepsilon_{b} /\varepsilon_{m} ) }\kappa \ell_s (1+\kappa\lambda)\big]^2
}{\varepsilon_b C_\perp \over \varepsilon_\perp C_z} < 1,
\label{criterion}\end{align}
where $\ell_s$ is screening length of the metal and $\varepsilon_m$ its relative permittivity. As we can see, the critical screening for single-domain ferroelectricity depends on both bulk and interface parameters in a non-trivial way. In any case, the more realistic condition \eqref{criterion} is less restrictive than \eqref{condition-on-gradients} due to the non-idealness of the metals.

We note that effective stiffness of the ferroelectric with respect to space variations is determined by both the gradient coefficients $C_z$ and $C_\perp$ and the permittivities $\varepsilon_b $ and $ \varepsilon_\perp$ as they enter in \eqref{criterion}. In fact, the ratio $\varepsilon_b / \varepsilon_\perp $ acts as an effective anisotropy that suffices, by itself, to activate the appearance of multi-domain ferroelectricity due to interfacial bonding effects. 
In PbTiO$_3$/SrRuO$_3$/SrTiO$_3$ capacitors, for example, $\varepsilon_b / \varepsilon_\perp \sim 10^{-2}$ \cite{Pertsev98}
and the condition \eqref{criterion} is fulfilled for interfaces with $\lambda = 0$. Consequently, irrespective of the screening strength of the metals, a multi-domain transition has to be expected in this case.
In the case of BaTiO$_3$, in addition, there is a difference between $C_z$ and $C_\perp$ and the ratio ${\varepsilon_b C_\perp\over \varepsilon_\perp C_z}\sim 10^{-2}$. It can be estimated that, to make it possible a single-domain transition, the interface energy needs to be $|\lambda| \gtrsim 1$ nm in both type of capacitors, which further restrics the maximum possible enhancement of ferroelectricity.

In summary, we have studied the role that interfacial bonding plays in the paraelectric instability of ferroelectric capacitors. Bonding effects are shown to have an unexpected impact on the competition between single- and multi-domain ferroelectricity, and therefore turn out to be crucial for determining the actual ground state of the system. 
The reason behind this is the intrinsic depolarizing field that appear inside the ferroelectric when the polarization varies in space in order to accommodate to the interfaces. 
We have found that there exists a threshold in the anisotropy of the ferroelectric beyond which the ground state correspond to multi-domain ferroelectricity. This threshold has been determined analitically in terms of the material parameters that characterize the ferroelectric and its interfaces (that is, gradient coefficients, non-ferroelectric permittivities and extrapolation length). 
Thus we have shown that the critical screening strength necessary to keep the system in its single-domain state is unexpectedly entangled with the interface properties. These results deepen our understanding of the fundamental properties of ferroelectrics at the nanoscale and are expected to motivate further studies on the interplay between ferroelectricity and interfacial phenomena. 

We thank P. Bruno, E. Kats, and B. Misirlioglu for useful discussions. 
A.P.L. was partially supported by Russian Foundation
of Basic Research grant \# 13-02-12450-ofi-m.

\newpage
\
\newpage
\onecolumngrid

\appendix 

{\centering
{\large \sc Supporting Information}

\vspace{10pt}

\setcounter{page}{1}
\setcounter{equation}{0}

{ \large \bf Single- and multi-domain ferroelectricity driven by metal-oxide bonding}

\vspace{20pt}
A. Cano$^{1,2}$ and A.P. Levanyuk$^3$

\small \it 
$^1$CNRS, Univ. Bordeaux, ICMCB, UPR 9048, F-33600 Pessac, France
\mbox{$^2$European Synchrotron Radiation Facility, 6 rue Jules Horowitz, BP 220, 38043 Grenoble, France
}
\mbox{$^3$Departamento de Física de la Materia Condensada, C-III, Universidad Autónoma de Madrid, E-28049 Madrid, Spain}

}

\

The Landau-like approach was first employed for the systematic study of the ferroelectric instabiltiy in thin films in \cite{Chensky82}, and since then it has been very helpful for understanding many aspects of the physics behind this phenomenon. The state-of-the-art of this method is reviewed in \cite{Bratkovsky09}.  
In the following we outline the generalization developed in our work, indicating explicitly the ingredients necessary to incorporate interfacial effects and get our results.  

For the sake of clarity, we first consider the case of ideal electrodes. Then the ferroelectric is described by means of the equations \eqref{constitutive} and \eqref{Gauss} under the boundary conditions \eqref{EBC} and \eqref{ABC}:
\begin{align}
[A - C_\perp (\partial_x^2 +\partial_y^2 )- C_z \partial_z^2]P &= - \partial_z V,
\tag{S.1} 
\label{constitutive-sup}\\
\varepsilon_0\varepsilon_\perp (\partial_x^2 +\partial_y^2 )V + \varepsilon_0\varepsilon_b \partial_z^2 V  - \partial_z P &= 0,
\tag{S.2} 
\label{Gauss-sup}
\end{align}
with
\begin{align}
V(z = \pm l/2) =0, \qquad 
\left.(1 \pm \lambda \partial_z ) P\right|_{z = \pm l/2} = 0. 
\tag{S.3} 
\end{align}
Generally, in the first equation, the derivatives of $P$ with respect to $z$ are neglected compared to the rest of terms:
\begin{align}
[A - C_\perp (\partial_x^2 +\partial_y^2 )- \xcancel{C_z \partial_z^2}]P = - \partial_z V.
\tag{S.4} 
\end{align}
In doing so, one effectively ignores the boundary conditions for the polarization or, more precisely, one tacitly assumes that they correspond to the natural boundary conditions $\left. \partial_z P \right|_{z = \pm l/2}=0 $ associated to passive interfaces. To take into account interfacial effects, we have to keep all the terms in the above equations and proceed as follows. 

The form of these equations is such that the polarization can be expanded in a Fourier series:
\begin{align}
P(\mathbf r) = \sum_{k_x,k_y} p_{k_x,k_y}(z)\cos k_x x \cos k_y y
\tag{S.5} 
\end{align}
and similarly for $V (\propto P)$. In the paraelectric phase $P = 0 $, while $P \not = 0 $ when this phase becomes unstable and the systems enters in the ferroelectric phase. Since the equations are linear, this happens due to one of the terms in the Fourier series (which will act as a source for the rest of terms in the series through nonlinear terms in the equations). Thus, to determine the transition point, the task is to find out the term in the Fourier series that gives the first non trivial solution of the linear equations. The best candidates are the 2D distributions
\begin{align}
P=p_{k}(z)\cos k x,
\qquad \text{ or similarly with } x \leftrightarrow y ,
\tag{S.6} 
\end{align}
because, compared to full 3D distributions ($k_x, k_y \not =0$), the gradient energy is minimized. Substituting in \eqref{constitutive-sup} and \eqref{Gauss-sup} we then get the ordinary differential equations:
\begin{align}
(A + C_\perp k ^2)p - C_z p'' &= - v',
\tag{S.7} 
\\
-\varepsilon_0\varepsilon_\perp k^2 v + \varepsilon_0\varepsilon_b v'' - p' &= 0,
\tag{S.8} 
\end{align}
for $p$ and $v$. Taking into account the symmetry of the system, the general solution of these equations is 
\begin{align}
p (z)&= p_1 \cos k_1 z + p_2 \cos k_2 z
\tag{S.9} 
\\
v (z)&= p_1 {k_1 \over \varepsilon_0 \varepsilon_\perp k^2 + \varepsilon_0 k_1^2}
\sin k_1 z + p_2 {k_2 \over \varepsilon_0 \varepsilon_\perp k^2 + \varepsilon_0 k_2^2}\sin k_2 z
\tag{S.10} 
\end{align}
which corresponds to the distributions considered in the main text. Compared to previous considerations (see \cite{Bratkovsky09}), here we deal with two different functions to describe the $z$ dependence of the polarization. This is essential to take into account both electrostatic and interfacial boundary conditions simultaneously. By direct substitution in the equations and boundary conditions we obtain, after some trivial manipulations, the algebraic equations
\begin{gather}
\left(  \varepsilon_{0}A +\frac{C_{\perp}}{C_{z}}\frac{k^{2}%
}{\kappa_{0}^{2}}\right)  \varepsilon_{\perp}\frac{k^{2}}{\kappa_{0}^{2}%
}-\frac{k_{1}^{2}k_{2}^{2}}{\kappa_{0}^{4}} =0,
\label{al1}
\tag{S.11} 
\\
1+\varepsilon_{0}A +\frac{C_{\perp}+\varepsilon_{\perp}C_{z}}{C_{z}}%
\frac{k^{2}}{\kappa_{0}^{2}}+\frac{k_{1}^{2}+k_{2}^{2}}{\kappa_{0}^{2}}
=0,
\label{al2}
\tag{S.12} 
\\
\left(  \varepsilon_{0}A +\frac{C_{\perp}}{C_{z}}\frac{k^{2}%
}{\kappa_{0}^{2}}+\frac{k_{1}^{2}}{\kappa_{0}^{2}}\right)  \tan\frac{k_{1}%
l}{2}\left(  1-k_{2}\lambda\tan\frac{k_{2}l}{2}\right)  \frac{k_{2}}%
{\kappa_{0}}-\left(  \varepsilon_{0}A +\frac{C_{\perp}}{C_{z}}%
\frac{k_{y}^{2}}{\kappa_{0}^{2}}+\frac{k_{2}^{2}}{\kappa_{0}^{2}}\right)
\tan\frac{k_{2}l}{2}\left(  1-k_{1}\lambda\tan\frac{k_{1}l}{2}\right)
\frac{k_{1}}{\kappa_{0}} =0,
\label{al3}
\tag{S.13} 
\end{gather}
where $\kappa_{0}=(\varepsilon_{0}C_{z})^{-1/2}$. These equations have to be satisfied in order to get a non trivial solution ($p_{1,2} \not =0$). To obtain temperature at which this happens as a function of the wavenumber $k$ (plotted in Fig. \ref{F2}), the parameters $k_1$ and $k_2$ can be expressed in terms of $k$ from these equations and further used to obtain the function $A(k)$. The maximum of this function gives us the transition temperature and the period of the structure that appears at this point. The function $A(k)$ is found to be
\begin{align}
A(k)
=
A_0 \left\{ 
1  + 
\left[ (1 + \kappa \lambda)^2{\varepsilon_b C_\perp \over \varepsilon_\perp C_z} - {1\over 3}\right]
\left({kl \over 2}\right)^2 + \mathcal O (k^4)%
\right\},
\tag{S.14} 
\end{align}
from which we obtain the criterion Eq. \eqref{condition-on-gradients}. As we can see, this function necessarily has its maximum for $k\not =0$ if the condition \eqref{condition-on-gradients} is fulfilled. In other words, the transition necessarily implies multi-domain ferroelectricity in that case.

The curves shown in Figs. \ref{F2}, \ref{F3} and \ref{F4} are obtained from the numerical solution of the above set of algebraic equations [Eqs. \eqref{al1}, \eqref{al2} and \eqref{al3}].

In the case of electrodes with finite screening strengh, within a semi-classical approximation the electric potential for $|z|>l/2$ is such that 
\begin{align}
(\nabla ^2 + \ell_s^{-2})V =0,
\tag{S.15} 
\end{align}
with the condition that $V(z \to \pm \infty) =0$. Here $\ell_s$ represents the screening length. The analysis proceeds basically along the same lines, with the correspoding replacement of the electrostatic boundary condition at $z = \pm l/2$. 


\begin{references}
\bibitem{note_polar} The most general boundary condition reads $\left.(1 \pm \lambda \partial_z ) P\right|_{\Sigma} = \zeta$, where $\zeta$ accounts for the polar character of the $\Sigma$ interface. This term, however, plays no role in the paraelectric instability when the interfaces are equal as assumed in our case.
\end{references}
\end{document}